\documentclass{gMOS2e}

\usepackage{epstopdf}
\usepackage{subfigure}
\usepackage{upgreek}

\begin{document}


\articletype{GUIDE}

\title{\textit{Monte Carlo simulation of binary mixtures of hard colloidal cuboids}}

\author{
\name{A. Patti\textsuperscript{a}$^{\ast}$\thanks{$^\ast$Corresponding author. Email: alessandro.patti@manchester.ac.uk}
and A. Cuetos\textsuperscript{b}}
\affil{\textsuperscript{a}School of Chemical Engineering and Analytical Science, The University of Manchester, Manchester, M13 9PL, UK;
\textsuperscript{b}Department of Physical, Chemical and Natural Systems, Pablo de Olavide University, 41013 Sevilla, Spain}
\received{v4.0 released June 2015}
}

\maketitle

\begin{abstract}

We perform extensive Monte Carlo simulations to investigate the phase behaviour of colloidal suspensions of hard board-like particles (HBPs). While theories restricting particle orientation or ignoring higher ordered phases suggest the existence of a stable biaxial nematic phase, our recent simulation results on monodisperse systems indicate that this is not necessarily the case, even for particle shapes exactly in between prolate and oblate geometries, usually referred to as self-dual shape. Motivated by the potentially striking impact of incorporating biaxial ordering into display applications, we extend our investigation to bidisperse mixtures of short and long HBPs and analyse whether size dispersity can further enrich the phase behaviour of HBPs, eventually destabilise positionally ordered phases and thus favour the formation of the biaxial nematic phase. Not only do our results indicate that bidisperse mixtures of self-dual shaped HBPs cannot self-assemble into biaxial nematic phases, but they also show that these particles are not able to form uniaxial nematic phases either. This surprising behaviour is also observed in monodisperse systems. Additionally, bidisperse HBPs tend to phase separate in coexisting isotropic and smectic phases or, at relatively large pressures, in a smectic phase of mostly short HBPs and a smectic phase of mostly long HBPs. We conclude that limiting the particle orientational degrees of freedom or neglecting the presence of positionally ordered (smectic, columnar and crystal) phases can dramatically alter the phase behaviour of HBPs and unrealistically enlarge the region of stability of the biaxial nematic phase.

\end{abstract}


\section{Introduction}

Colloids are dispersions of solid particles or liquid droplets, between 1 nm and 1 $\upmu$m in size, evenly suspended in a fluid. Their motion, basically controlled by the stochastic collisions with the fluid molecules, is driven by a thermal energy of the order of few $kT$ per particle. Very interestingly, the interparticle forces controlling their phase behaviour and dynamics, including electrostatic and excluded volume interactions, are the same forces that determine the behaviour of atoms and molecules. As such, the possibility to synthesise particles in a variety of sizes and shapes and control their interactions, makes colloids ideal model systems to understand the physical laws underpinning structure and dynamics of atomic and molecular systems. This remarkable similarity provides a unique opportunity to unravel a number of processes at the molecular scale that are too fast to be detected by conventional microscopy. In particular, investigating how anisotropic colloidal particles organise themselves can contribute to unveil the phase and aggregation behaviour of molecular liquid crystals (LCs), enhance their fundamental understanding and thus optimise the design of functional electro-optical devices \cite{veinot}. Nevertheless, the scientific relevance of colloids goes well beyond their use as mere model systems. The impressive advances in the synthesis of anisotropic colloidal particles with precise symmetry and directional interactions sparked the discovery of a collective behaviour, key for the synthesis of photonic crystals \cite{li} and macroporous solids \cite{mourad}, that is not observed in atomic and molecular systems. Recognising this breakthrough has conferred to colloids a position in materials science in their own right \cite{manoharan}. 

Although current LC display technology is entirely based on molecular LCs, the appealing scenario of employing materials with high thermal stability, enhanced susceptibility to external fields and more accessible production costs, makes colloidal LCs excellent candidates for displays \cite{gabriel, lekkerkerker, anke}. Additionally, and perhaps more interestingly, colloidal suspensions of board-like particles can form biaxial nematic (N$_\text{B}$) phases \cite{vandenpol}, whose existence, theoretically predicted by Freiser almost 50 years ago \cite{freiser}, is still an open question at the molecular scale. The intriguing prospect of manufacturing biaxial LCDs has been enfeebled by the difficulty of obtaining a stable molecular N$_\text{B}$ phase, especially at convenient temperatures for display applications. The experimental findings by Vroege and coworkers, who observed a remarkably stable N$_\text{B}$ phase in systems of polydisperse goethite particles, have provided renewed expectations and perhaps an indication on how, at the molecular scale, the stability of the N$_\text{B}$ phase might be enhanced \cite{vandenpol}. 

Our recent work on a wide range of oblate and prolate monodisperse HBPs highlighted the existence of a rich variety of LC phases, including the long-debated discotic smectic (Sm) phase, consisting of layers as thick as the particle minor axis \cite{cuetos1}. However, no evidence of the existence of the N$_\text{B}$ phase could be provided, even at the so-called self-dual shape, a particle geometry almost exactly in between prolate and oblate. This peculiar particle shape was shown to favour the formation of the N$_\text{B}$ phase, although into a very limited region of the phase diagram \cite{taylor}. It is important to note that most of the theoretical studies on board-like particles were developed within the restricted-orientation (Zwanzig) model, allowing only six orthogonal particle orientations \cite{zwanzig, belli1, taylor}, or assuming complete alignment of the particle long axes \cite{vanakaras}. The seminal works by Straley \cite{straley} and Mulder \cite{mulder} on hard sphero-platelets (similar, but not identical, to our HBPs) did incorporate free rotation, but neglected the formation of positionally ordered phases, which were later shown to dramatically reduce the region of stability of the N$_\text{B}$ phase and deeply change the resulting phase diagram \cite{taylor}.
By applying a fundamental-measure theory within the Zwanzig approximation, Velasco and coworkers investigated the phase behaviour of HBPs as a function of the degree of particle biaxiality, defined as $\theta=(L^*-1)^{-1} \left( \frac{L^*}{W^*}-W^*\right)$, where $L^*$ and $W^*$ are the reduced particle length and width, respectively \cite{velasco}. In particular, $\theta=-1$ for oblate geometries or $\theta=1$ for prolate geometries. They observed a transition from the uniaxial nematic (N$_\text{U}$) to the N$_\text{B}$ phase for $\theta \lesssim 0$, but to the Sm phase for $\theta \gtrsim 0$. However, in our recent work on freely rotating monodisperse HBPs within this range of particle biaxiality ($\theta=-0.0909$), we only observed a transition to the Sm phase \cite{cuetos1}. As unambiguously stressed by Masters, overlooking the formation of ordered phases could be perilous for a reliable theoretical description of the N$_\text{B}$ phase behaviour \cite{masters}. We add that employing a restricted-orientation model can also have a similar effect, as our theoretical predictions on freely rotating HBPs have recently established \cite{cuetos1}. Monte Carlo simulations on hard spheroplatelets showed the existence of N$_\text{B}$ phases only for particle length-to-thickness ratios $L^*>9$ and dimensions close to the self-dual shape \cite{peroukidis}. Since our results on sharp HBPs with $L^*=12$ did not reveal the existence of the N$_\text{B}$ phase \cite{cuetos1}, we can only argue that particle roundness might play a crucial role in its stabilisation.

In the light of the conclusions drawn from our work on monodisperse HBPs, here we investigate the phase behaviour of binary mixtures of HBPs with a twofold aim: (\textit{i}) determining what LC phases are formed and, specifically, the region of existence, if any, of the N$_\text{B}$ phase; and (\textit{ii}) pondering the validity of restricting the particle rotational degrees of freedom to describe their self-assembly. To this end, the geometry of the HBPs in the present contribution is exactly the same as that employed by Belli \textit{et al} in their theoretical work on bidisperse and polydisperse HBPs within the Zwanzig model \cite{belli1}.


\section{Methodology}

\subsection{Model}
In Figure 1, we provide a visual representation of the particles studied in this work. They are freely-rotating hard rectangular parallelepipeds (cuboids) of thickness $T$, width $W$, and length $L$. The thickness $T$ is the unit length, while the reduced width and length are, respectively, $W^*=W/T$ and $L^*=L/T$. Following the theoretical work by Belli \textit{et al}, we make use of a geometrical parameter, $s$, to describe the degree of particle size dispersity \cite{belli1}. The dimensions of the short and long HBPs are then defined as a function of this parameter. For long HBPs, thickness, width and length are, respectively, $T_1=T(1+s)$, $W_1=W(1+s)$ and $L_1=L(1+s)$. Similarly, for short HBPs, the dimensions are $T_2=T(1-s)$, $W_2=W(1-s)$ and $L_2=L(1-s)$. By definition, the reduced length and width of the two species are the same: $L^*=L_1/T_1=L_2/T_2=L/T$ and $W^*=W_1/T_1=W_2/T_2=W/T$. In order to compare our simulation results with the above mentioned theoretical predictions, we assign to these two aspect ratios the same values as those in Ref. \cite{belli1}, namely $L^*=9.07$ and $W^*=2.96$. This particular geometry was chosen to reproduce the experimental observations by Vroege and coworkers, who found a remarkably stable N$_\text{B}$ phase in systems of polydisperse boardlike particles \cite{vandenpol}. Finally, the size-dispersity index is set to  $s=0.2$, a value that provides the richest phase behaviour among the binary mixtures investigated in Ref. \cite{belli1}. While our particles are identical to those designed by Belli, they are completely free to assume any possible orientation, rather than only the six allowed by the Zwanzig model. This difference is particularly relevant when assessing the formation of uniaxial and biaxial phases, as we have recently noticed in systems of monodisperse boardlike particles \cite{cuetos1}.

\begin{figure}[ht!]
\centering
  \includegraphics[width=0.7\columnwidth]{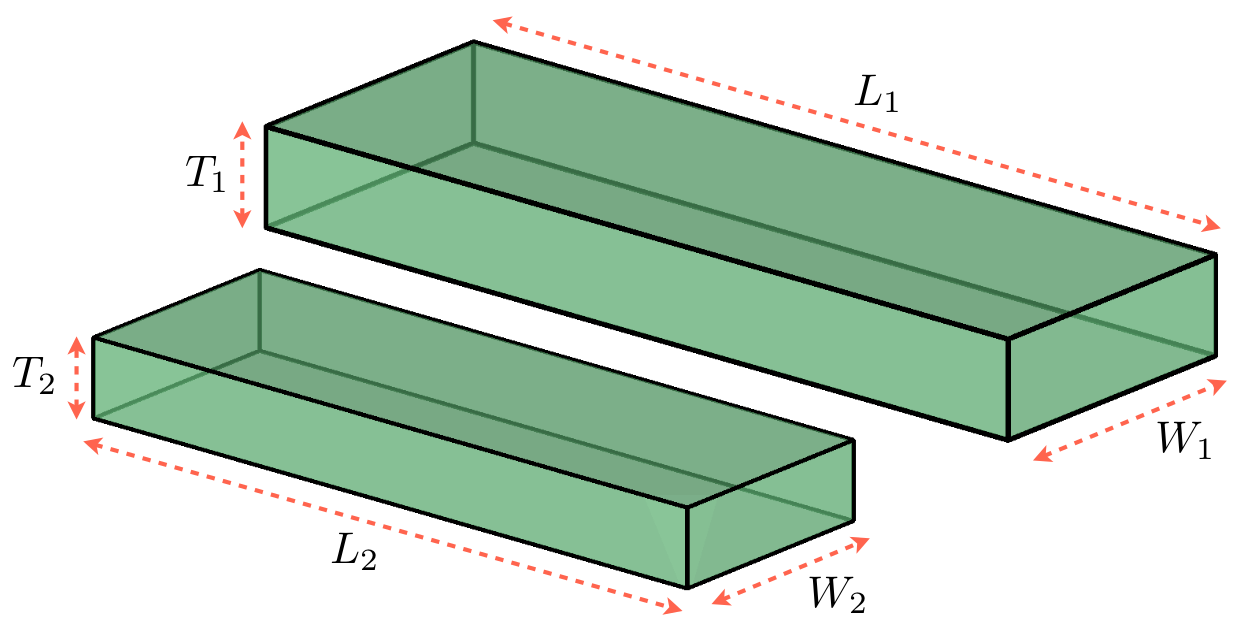}
  \caption{Long and short board-like particles with thickness, width and length given by ($T_1$, $W_1$, $L_1$) and ($T_2$, $W_2$, $L_2$), respectively. The reduced length $L^*=L_1/T_1=L_2/T_2$ and width $W^*=W_1/T_1=W_2/T_2$ are the same for both species. The size-dispersity index of the particles shown in this figure is $s=0.2$ (see text).}
\end{figure}

\subsection{Monte Carlo simulations}
We perform Monte Carlo simulations of systems containing between $N=2000$ and $N=4000$ HBPs, where $N=N_1+N_2$ is the sum of long ($N_1$) and short ($N_2$) HBPs. Our cuboids interact via a hard-core potential and, as such, their path towards the equilibrium state, which only depends on their geometry and system packing, can be investigated in terms of free volume available, or, equivalently, by checking and discarding the occurrence of particle overlaps. To this end, we applied the separating axes method described by Gottschalk \textit{et al} \cite{gottschalk} and adapted by John and Escobedo to study the phase behaviour of tetragonal parallelepiped particles \cite{john1}. We refer the interested reader to the Appendix of Ref. \cite{john1} for additional details. 

All the simulations were run in the isobaric-isothermal ensemble, which constrains a constant number of HBPs into a periodic simulation box of volume $V=L_x \times L_y \times L_z$, with $L_x$, $L_y$, and $L_z$ the box dimensions. To study phase coexistence, we decided to apply a direct coexistence simulation method, where the two species are initially separated by an interface and thus occupy different regions of the simulation box. In both regions, the particles are perfectly aligned, so to create a dense biaxial nematic phase consisting of short or long HBPs. Undesired finite size effects have been taken into account by employing elongated boxes, where $L_x$ is at least four or five times larger than the other two dimensions. This choice ensures a bulk behaviour in the regions far from the interface and allows us to impose an isotropic pressure gradient to a system that, due to the presence of an interface, is actually affected by an anisotropic pressure tensor. However, as long as the box is sufficiently elongated, this pressure tensor, scaling with $L_x^{-1}$, can be safely neglected \cite{noya}. The systems in such an initial configuration, containing the two species at different relative concentrations, in the range $0 \leq x_1=N_1/N \leq 1$, have been gradually expanded or compressed to the target reduced pressure $P^*=\beta PT^3$, where $\beta$ is the inverse temperature. Moreover, to better identify the boundary of the isotropic monophasic regions, we also compressed systems of randomly mixed HBPs. This also allowed us to confirm that the interface imposed in our direct coexistence simulations was actually the one with the lowest free energy. Each MC cycle consisted of $N$ attempts of displacing and/or rotating randomly selected particles, plus an attempt to modify the three box lengths independently. Translational and rotational moves as well as volume changes were accepted if no overlap was detected. 

To address equilibration, we calculated the uniaxial ($S_2$) and biaxial ($B_2$) order parameters as well as the packing fraction $\eta=\sum(N_jv_j)/V$, with $v_j=T_j \times W_j \times L_j$ the volume of a particle of species $j=1$ or 2. The systems were considered to be at equilibrium when $\eta$, $S_2$ and $B_2$ achieved a steady value within reasonable statistical fluctuations. To weight the contribution of each species to the global order of the system, the order parameters have been calculated for short and long HBPs separately. In particular, to determine the nematic order parameter and nematic director associated to each particle axis, the following traceless symmetric second-rank tensor has been diagonalized:

\begin{equation}
{\bf Q}^{\lambda \lambda}= \frac{1}{2N} \left\langle \, \sum^{N}_{i=1} (3\pmb{{\hat{\lambda}}}_{i} \cdot \pmb{{\hat{\lambda}}}_{i}
\,-\,{\bf I})\, \right\rangle,
\end{equation}

\noindent where $\pmb{{\hat{\lambda}}}$=$\bf{{\hat{x}}}$, $\bf{{\hat{y}}}$, $\bf{{\hat{z}}}$ refers to the unit orientation vectors of the generic particle $i$ along $W$, $T$,and $L$, respectively, ${\bf I}$ is the second-rank unit tensor, and the angular brackets indicate ensemble average. Three eigenvalues and three eigenvectors are obtained from diagonalization of ${\bf Q}^{\lambda \lambda}$, with the largest positive eigenvalue and the associated eigenvector determining, respectively, the nematic order parameter $S_{2}$ and the uniaxial nematic director \cite{eppenga}. In particular, the nematic order parameter $S_{2,L}$ is the largest eigenvalue of the tensor ${\bf Q}^{zz}$ and reads 

\begin{equation}
 S_{2,L}=\hat{{\bf n}}\cdot{\bf Q}^{zz} \cdot \hat{{\bf n}},
\end{equation}

\noindent where the eigenvector $\bf{\hat{n}}$ is the nematic director associated to the preferential orientation of the particle axis $\bf{{\hat{z}}}$. Similar expressions can be obtained to determine the remaining two uniaxial order parameters, $S_{2,W}=\hat{{\bf m}}\cdot{\bf Q}^{xx} \cdot \hat{{\bf m}}$ and $S_{2,T}=\hat{{\bf l}}\cdot{\bf Q}^{yy} \cdot \hat{{\bf l}}$, with $\hat{{\bf m}}$ and $\hat{{\bf l}}$ their respective nematic directors. By contrast, the biaxial order parameter associated to the nematic director $\bf{\hat{n}}$ is calculated as follows \cite{camp} 

\begin{equation}\label{eq2}
B_{2,L}= \frac{1}{3} \left( \hat{{\bf m}}\cdot{\bf Q}^{xx}\cdot \hat{{\bf m}}+\hat{{\bf l}}\cdot{\bf Q}^{yy} \cdot \hat{{\bf l}}-
\hat{{\bf m}} \cdot {\bf Q}^{yy} \cdot \hat{{\bf m}} - \hat{{\bf l}} \cdot {\bf Q}^{xx} \cdot \hat{{\bf l}}  \right).
\end{equation}

\noindent In principle, $B_{2,W}$ and $B_{2,T}$ can be obtained by similar expressions. However, to assess phase biaxiality is enough to monitor the biaxial order parameter associated to the principal particles axis, defined as the axis displaying the largest uniaxial order parameter \cite{allen, camp2, teixeira}. Therefore, here we only consider the biaxial order parameter associated to the main nematic director and refer to it as $B_2$. Additional details are available in our recent work on monodisperse HBPs \cite{cuetos1}. Finally, to calculate the particle composition in each coexisting phase, we run $NVT$ simulations within the miscibility gap at $x_1=0.5$ and, in some cases to improve statistics, along the same tie-line at $x_1=0.2$, 0.4, 0.6 and 0.8. We then measured the particle composition by dividing the box in layers of cross section $A_c=L_y \times L_z$ and height $L_1$ parallel to $L_x$. 


\section{Results}
Details of the binary mixtures investigated in this work are reported in Table 1, where we provide a summary of the ordering observed in the monophasic and biphasic regions. The $P^* - x_1$ (pressure-composition) phase diagram of these systems is given in Figure 2.

\begin{table}[h!]
\tbl{Details of the systems studied in this paper, consisting of binary mixtures of short and long HBPs. For comparison, we report the reduced pressure $P^{*}$, phases at equilibrium\textsuperscript{a}, composition of long HBPs $x_1$, packing fraction $\eta$, and uniaxial and biaxial order parameters. Superscripts $^{(1)}$ and $^{(2)}$ refer to the phases at equilibrium in the biphasic region of the phase diagram.}
{\begin{tabular}  {@{}c|cc|cc|cc|cc|cc}
\toprule
$P^*$ & Phase 1 & Phase 2 & $x_1^{(1)}$ & $x_1^{(2)}$ & $\eta^{(1)}$ & $\eta^{(2)}$ & $S_{2,L}^{(1)}$ & $S_{2,L}^{(2)}$ & $B_2^{(1)}$ & $B_2^{(2)}$ \\
\colrule
0.093 & \multicolumn{2}{c|}{I}	& \multicolumn{2}{c|}{1.0} & \multicolumn{2}{c|}{0.35} & 0.07 & -    & 0.02 & -    \\
0.094 & \multicolumn{2}{c|}{Sm}	& \multicolumn{2}{c|}{1.0} & \multicolumn{2}{c|}{0.39} & 0.79 & -    & 0.02 & -    \\
0.100 & \multicolumn{2}{c|}{I}	& \multicolumn{2}{c|}{0.5} & \multicolumn{2}{c|}{0.31} & 0.05 & 0.02 & 0.02 & 0.02 \\
0.108 & \multicolumn{2}{c|}{I}	& \multicolumn{2}{c|}{0.8} & \multicolumn{2}{c|}{0.36} & 0.08 & 0.05 & 0.02 & 0.02 \\
0.115 & \multicolumn{2}{c|}{I}	& \multicolumn{2}{c|}{0.4} & \multicolumn{2}{c|}{0.35} & 0.12 & 0.05 & 0.02 & 0.02 \\
0.125 & \multicolumn{2}{c|}{I}	& \multicolumn{2}{c|}{0.1} & \multicolumn{2}{c|}{0.26} & 0.08 & 0.03 & 0.04 & 0.02 \\
0.320 & \multicolumn{2}{c|}{I}	& \multicolumn{2}{c|}{0.0} & \multicolumn{2}{c|}{0.36} & -    & 0.05 & -    & 0.02 \\
0.335 & \multicolumn{2}{c|}{Sm}	& \multicolumn{2}{c|}{0.0} & \multicolumn{2}{c|}{0.42} & -    & 0.91 & -    & 0.02 \\
\colrule
0.125 & Sm   & I & 0.86 & 0.33 & 0.50 & 0.30 & 0.61 & 0.08 & 0.06 & 0.01 \\
0.140 & Sm   & I & 0.88 & 0.26 & 0.49 & 0.30 & 0.61 & 0.06 & 0.05 & 0.01 \\
0.150 & Sm   & I & 0.90 & 0.17 & 0.57 & 0.24 & 0.74 & 0.09 & 0.11 & 0.02 \\
0.175 & Sm   & I & 0.93 & 0.07 & 0.55 & 0.26 & 0.72 & 0.04 & 0.22 & 0.03 \\
0.200 & Sm   & I & 0.92 & 0.05 & 0.56 & 0.27 & 0.78 & 0.09 & 0.22 & 0.01 \\
0.230 & Sm   & I & 0.93 & 0.04 & 0.63 & 0.30 & 0.78 & 0.08 & 0.22 & 0.02 \\
0.250 & Sm   & I & 0.93 & 0.04 & 0.60 & 0.28 & 0.76 & 0.04 & 0.26 & 0.03 \\
0.275 & Sm$_B$ & I & 0.94 & 0.01 & 0.66 & 0.31 & 0.77 & 0.09 & 0.39 & 0.03 \\
0.300 & Sm$_B$ & I & 0.95 & $2\times 10^{-3}$ & 0.64 & 0.35 & 0.85 & 0.12 & 0.59 & 0.09 \\
0.320 & Sm$_B$ & I & 0.94 & $3\times 10^{-4}$ & 0.65 & 0.40 & 0.85 & 0.16 & 0.60 & 0.08 \\
0.350 & Sm$_B$ & Sm & 0.96 & $3\times 10^{-3}$ & 0.72 & 0.42 & 0.85 & 0.85 & 0.76 & 0.05 \\
\botrule
\end{tabular}}
\tabnote{\textsuperscript{a} Sm$_B$ indicates smectic phases with a significant biaxial order.}
\end{table}

Four main regions can be identified: (\textit{i}) the region of stability of the I phase, where both species are well-mixed together across the complete range of compositions; (\textit{ii}) the region of stability of a Sm phase that incorporates mostly long HBPs, referred to as Sm$_1$; (\textit{iii}) a dark gray shaded area being the region of I/Sm$_1$ coexistence; and (\textit{iv}) a light gray shaded area indicating equilibrium highly ordered phases. The empty squares indicate overall system concentrations, set to $x_1=0.4$, 0.5 or 0.6. Regardless of these values, the resulting coexistence concentrations in the I and Sm phases are generally in good agreement, which is of the order of the error bars shown in Figure 2. The reason why we are showing an overall concentration rather than an other is only due to the fact that some Sm phases display a number of defects whose annihilation time is especially long. Although the presence of these defects does not seem to particularly influence the concentrations at coexistence, in Figure 2 we decided to include those overall concentrations referring to Sm phases whose defects have been (almost) completely annihilated within our simulation time. At increasing pressure, the systems undergo a transition from the I phase to a two-phase region where I phases rich in short HBPs coexist with Sm phases rich in long HBPs. At larger pressures, we observe a region where two Sm phases coexist: Sm$_1$, rich in long HBPs, and Sm$_2$, mostly composed by short HBPs. The pure-component systems, indicated by the vertical lines $x_1=0$ and $x_1=1$, experience transitions from the I phase to the Sm phase with no evidence of intermediate N$_\text{U}$ phases. This result is rather remarkable as theoretical works on spheroplatelets \cite{taylor, mulder} and HBPs \cite{belli1, cuetos1} did identify the presence of stable N$_\text{U}$ phases at particle geometries equal or close to the self-dual shape, where the relation $\sqrt{LT}=W$ between particle dimensions holds. Nevertheless, as we could establish in our work on monodisperse HBPs with $L^*=9$ and $L^*=12$, the region of stability of the N$_\text{U}$ phase for this particle shape is so small that pinning down its boundaries becomes very challenging, especially due to the weak first-order character of the I-to-N phase transition of this particular particle geometry.

\begin{figure}[ht!]
\centering
  \includegraphics[width=0.7\columnwidth]{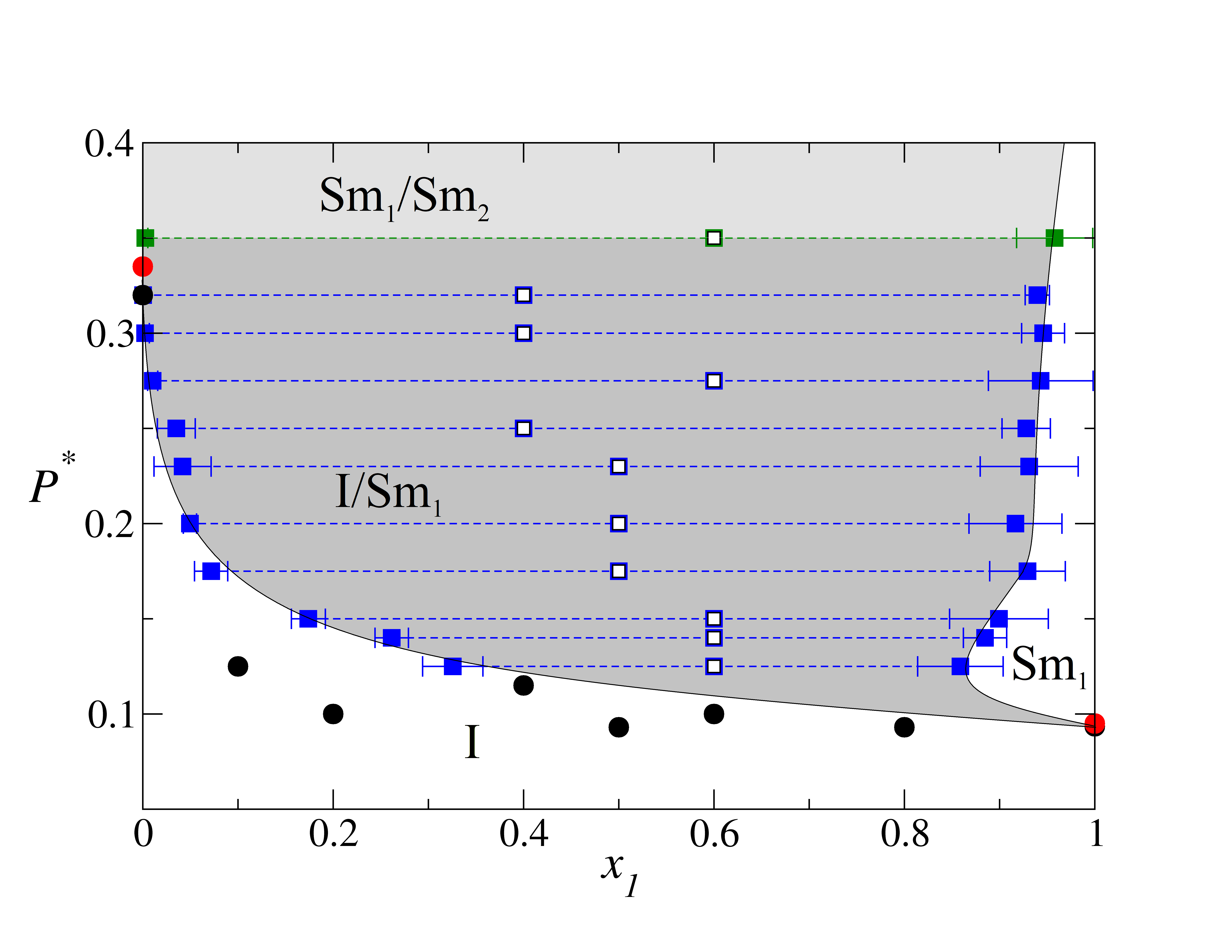}
  \caption{Pressure-composition phase diagram of HBPs with size-dispersity index $s=0.2$. Dark and light gray shaded areas represent the region of I/Sm$_1$ and Sm$_1$/Sm$_2$ coexistence, respectively. Empty squares indicate the overall system concentration of long HBPs in the biphasic regions, whereas solid squares, connected by dashed lines, indicate the concentration of long HBPs in each coexisting phase. Black and red circles represent pure isotropic (I) and smectic (Sm) phases, respectively. Error bars are standard deviations and solid lines are guides to the eye.}
\end{figure}

As a general tendency, the transition pressure from the monophasic to the biphasic region strictly depends on the relative content of short and long HBPs and spans from $P^*\approx0.328$ for $x_1=0.0$ to $P^*\approx0.094$ for $x_1=1.0$. In terms of packing fraction, the monophasic-to-biphasic phase transitions occur at approximately $0.35<\eta<0.40$, the same interval at which monodisperse self-dual-shaped HBPs \cite{cuetos1} and monodisperse hard spheroplatelets \cite{peroukidis} experience an I-to-N phase transition. Although the theory of Ref. \cite{belli1} does not predict the existence of an I/Sm biphasic region, it locates the I-to-N transition of binary mixtures of HBPs at $\eta \approx 0.28$, a packing fraction at which our binary mixtures are  still isotropic.

At pressures generally below $P^*<0.1$, depending on particle composition, a single I phase is found, where short and long particles are evenly mixed throughout the simulation box. Figure 3 displays a typical configuration of an I phase at $P^*=0.093$ and composition $x_1=0.8$. While the configuration in the top frame shows both particle species, that in the middle frame only consists of short HBPs to provide an unambiguous evidence of their homogeneous distribution thoughout the simulation box. We stress that the initial configuration of this mixture consisted of two separated phases, each incorporating either short or large HBPs. The average composition profiles along the longest box dimension for long and short HBPs are given in the bottom frame and confirm the presence of a well-mixed I phase. Although some small clusters of long HBPs are observed in the top frame, the long-ranged order of this system is very weak, with the pair correlation functions (not shown here) decaying fast to 1 at length scales comparable to $L_1$.

\begin{figure}[ht!]
\centering
  \includegraphics[width=0.7\columnwidth]{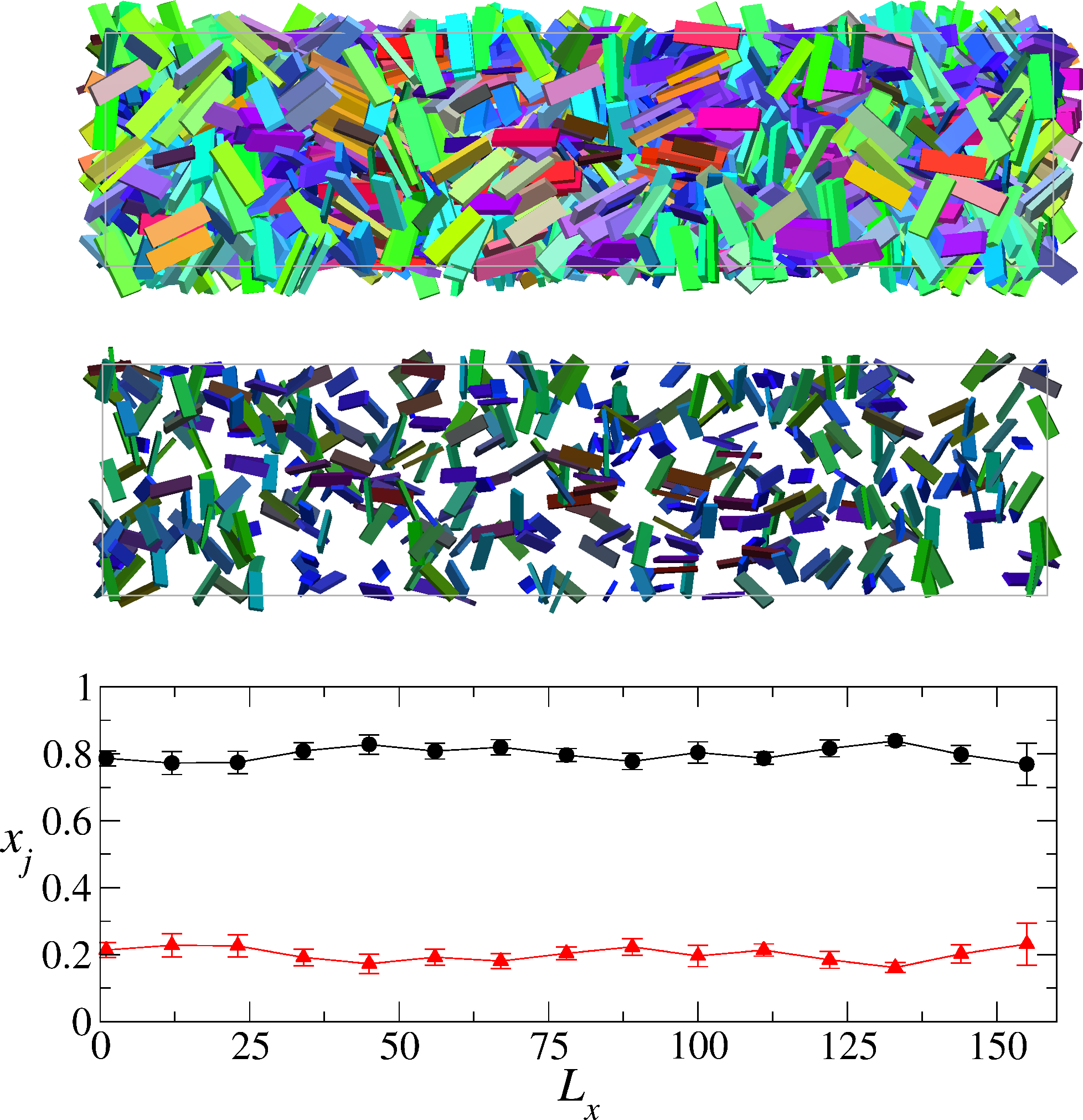}
  \caption{Snapshots of an isotropic phase at $P^*=0.093$ and $x_1=0.8$ containing $N=4000$ HBPs with size-dispersity index $s=0.2$. The snapshot in the top frame includes long and short HBPs, while that in the middle frame only short HBPs. The graph at the bottom reports the composition profile of long (black circles) and short (red triangles) HBPs along the longest box dimension. Error bars in the bottom frame indicate standard deviations and solid lines are guides to the eye.}
\end{figure}

The phase diagram in Figure 2 unveils the coexistence between a Sm$_1$ phase, rich in long HBPs, and an I phase, rich in short HBPs. A typical configuration of this intriguing phase coexistence is provided in Figure 4 for a binary system of 4000 HBPs at $P^*=0.13$ and $x_1=0.6$. The corresponding average composition profiles, provided in the bottom frame, unveil the formation of a Sm phase rich in long HBPs with $x_1\approx0.85$ (black circles) and an I phase rich in short HBPs with $x_1\approx0.28$ (red triangles). Interestingly enough, the former incorporates short HBPs in the interlayer spaces as well as within the smectic layers. Transverse interlayer particles had already been observed in Sm phases of monodisperse spherocylinders \cite{vanroij, allen1, allen2, Patti1, Patti2}, but their occurrence had always been considered particularly rare, especially for the prohibitive free-energy barriers to reach the transverse state \cite{allen2}. Here, it seems that the probability of observing transverse inter-layer HBPs is not so low, although we have not assessed this point quantitatively. In particular, short HBPs accommodate themselves in the spaces available between long HBPs and give rise to a very peculiar Sm phase with prolate-like layers, formed by short and long HBPs oriented along their main axis, and oblate-like layers, incorporating mainly short HBPs oriented along their minor axis. Transversely oriented long HBPs can in principle also be observed, but, as also found in monodisperse systems of spherocylidners, their occurrence is expected to be rare.

\begin{figure}[ht!]
\centering
  \includegraphics[width=0.7\columnwidth]{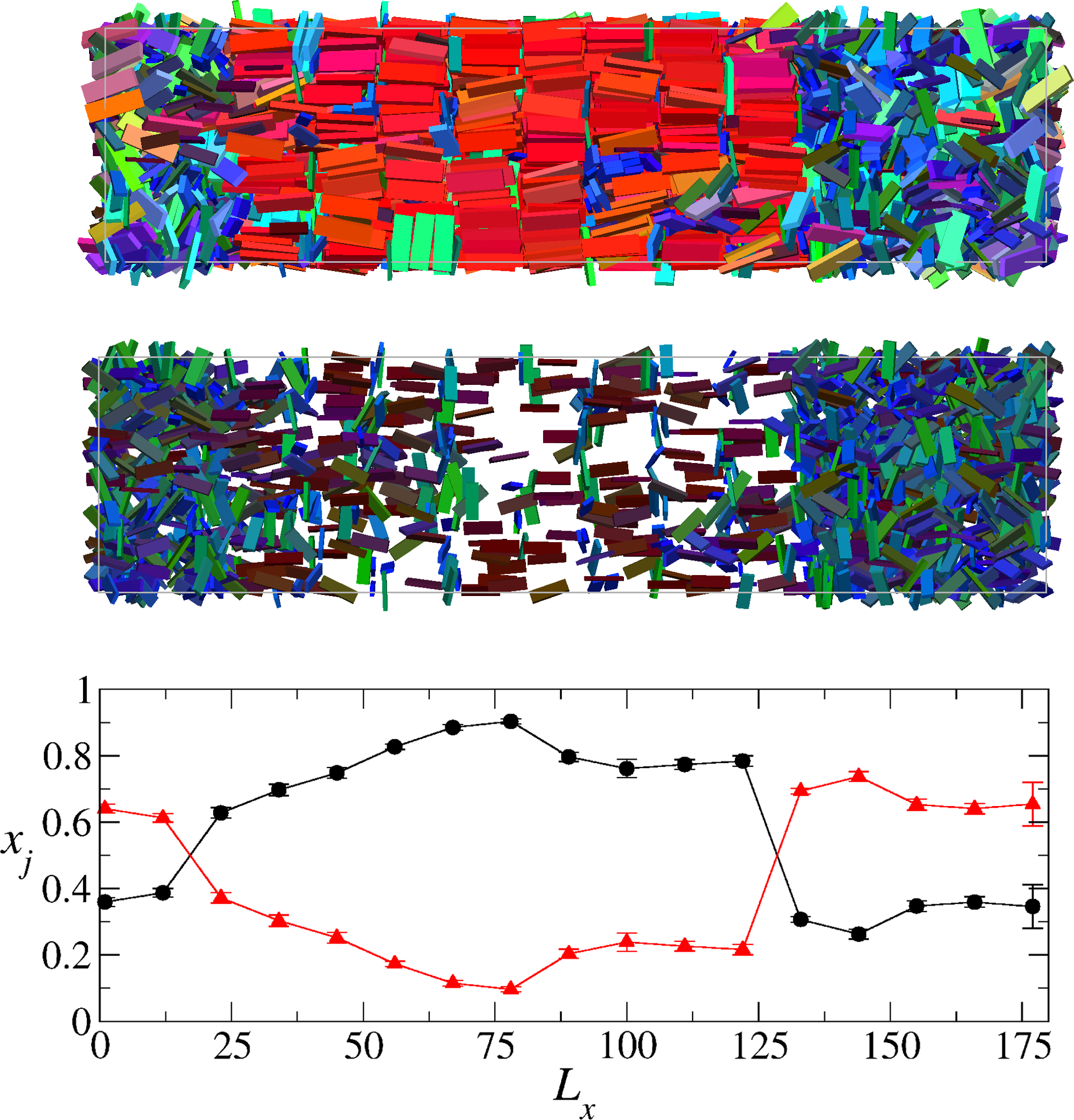}
  \caption{Snapshots of an isotropic phase rich in short HBPs at equilibrium with a smectic phase rich in long HBPs. Pressure, total number of particles, and global composition are, respectively, $P^*=0.13$, $N=4000$ and $x_1=0.6$. The configuration in the top frame includes long and short HBPs, while that in the middle frame only  short HBPs. The graph  at the bottom reports the composition profile of long (black circles) and short (red triangles) HBPs along the longest box dimension. Error bars in the bottom frame indicate standard deviations and solid lines are guides to the eye.}
\end{figure}

To quantify these observations, we calculated the longitudinal pair correlation function, $g_{\parallel}(\textbf{r}\cdot\bf{{\hat{n}}})$, between particles at distance $\textbf{r}$ projected along the nematic director $\bf{{\hat{n}}}$. It is convenient to calculate these functions in cubic rather than elongated boxes, where a single phase is observed. To this end, once that the boundaries of the miscibility gap had been determined, we equilibrated a number of Sm phases in the region of the $P^* - x_1$ phase diagram where monophasic equilibrium is expected. In Figure 5, we show $g_{\parallel}(\textbf{r}\cdot\bf{{\hat{n}}})$ for long and short HBPs in a pure Sm$_1$ phase at $P^*=0.15$ and $x_1=0.95$. The black solid line, $g_{11}$, suggests that the relative distance between long HBPs is $d/T\approx12$ in the direction of $\bf{{\hat{n}}}$, corresponding to a particle length ($L_1/T=10.884$) plus the interlayer spacing, approximately given by the short-particle thickness ($T_2/T=0.8$). The non-zero values of $g_{11}$ at intermediate distances between contiguous smectic layers confirm that long HBPs can also be observed in the interlayer region, where they can assume a transverse orientation. The red dashed line ($g_{22}$) provides the spatial correlations between short HBPs. Although in this case the peaks are significantly smaller than those of $g_{11}$, a well-defined pattern can still be recognised. While the main peaks of this pattern cannot unambiguously resolve whether short HBPs are located in the prolate-like or in the oblate-like layers, the minor peaks in between clearly indicate that they are most probably located in both of them. This conclusion is confirmed by the analysis of $g_{12}$, given by the green dot-dashed line in Figure 5, which describes the distribution of short HBPs with respect to long HBPs. The broadness of its main peaks denotes a relatively wide in-layer positional distribution of short HBPs, which, due to their reduced length as compared to the layer thickness, are free to fluctuate in the direction of $\bf{{\hat{n}}}$ and around the center of mass of the layers. The secondary peak at exactly $d/2T$ indicates that short HBPs are indeed laying in the interlayer spacing or, equivalently, forming oblate-like smectic layers.

\begin{figure}[ht!]
\centering
  \includegraphics[width=0.7\columnwidth]{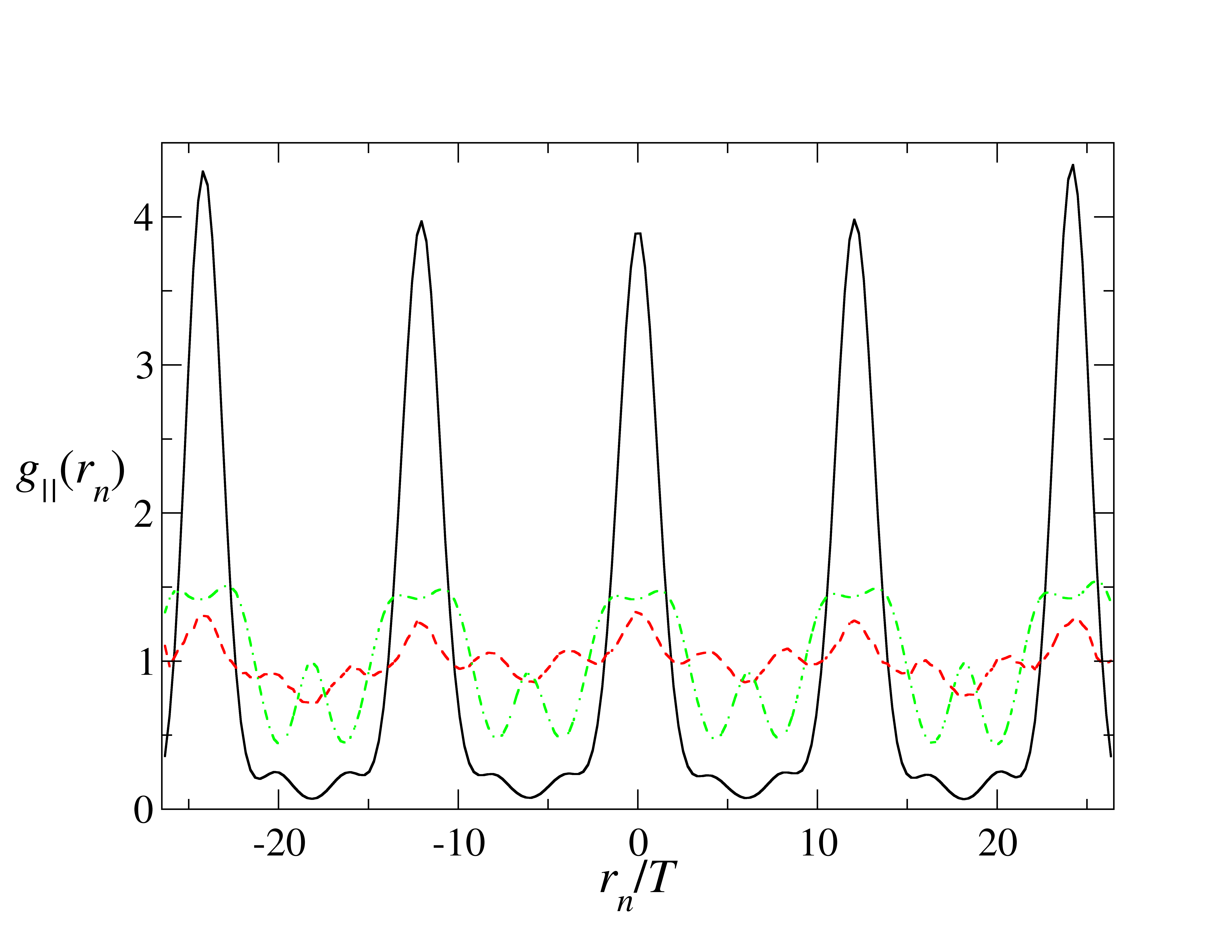}
  \caption{Longitudinal pair correlation function, $g_{\parallel}(r_n)$, with $r_n=\textbf{r}\cdot\bf{{\hat{n}}}$, of a pure Sm$_1$ phase along the direction of the nematic director $\bf{{\hat{n}}}$ at $x_1=0.95$ and $P^*=0.15$. Solid black, dashed red and dot-dashed green lines refer to the spatial correlations of long-long, short-short and long-short HBPs, respectively.}
\end{figure}

At larger pressures, above $P^*=0.32$, two coexisting Sm phases, one rich in long HBPs (Sm$_1$) and the other in short HBPs (Sm$_2$), are observed. An example of this two-phase equilibrium is shown in Figure 6, where 2000 cuboids at $P^*=0.35$ and $x_1=0.2$ give rise to two separate layered structures. Both phases present a number of unresolved defects, which are expected to be annihilated at time scales larger than our simulation time. While short HBPs are again observed in the bulk of Sm$_1$, it is very difficult to see long HBPs in the bulk of Sm$_2$, as the average composition profile in the bottom frame of Figure 6 points out. A simple visual inspection of the top frame of this figure suggests that the minor axes of long HBPs appear to be significantly aligned and thus that the Sm$_1$ phase might possess a degree of biaxiality. These observations are actually not limited to the region of the phase diagram where two Sm phases coexist, but also at lower pressures, where an I/Sm coexistence has been detected. To address this point, we have calculated the order parameter defined in Eq. (2) and (3). 

\begin{figure}[ht!]
\centering
  \includegraphics[width=0.7\columnwidth]{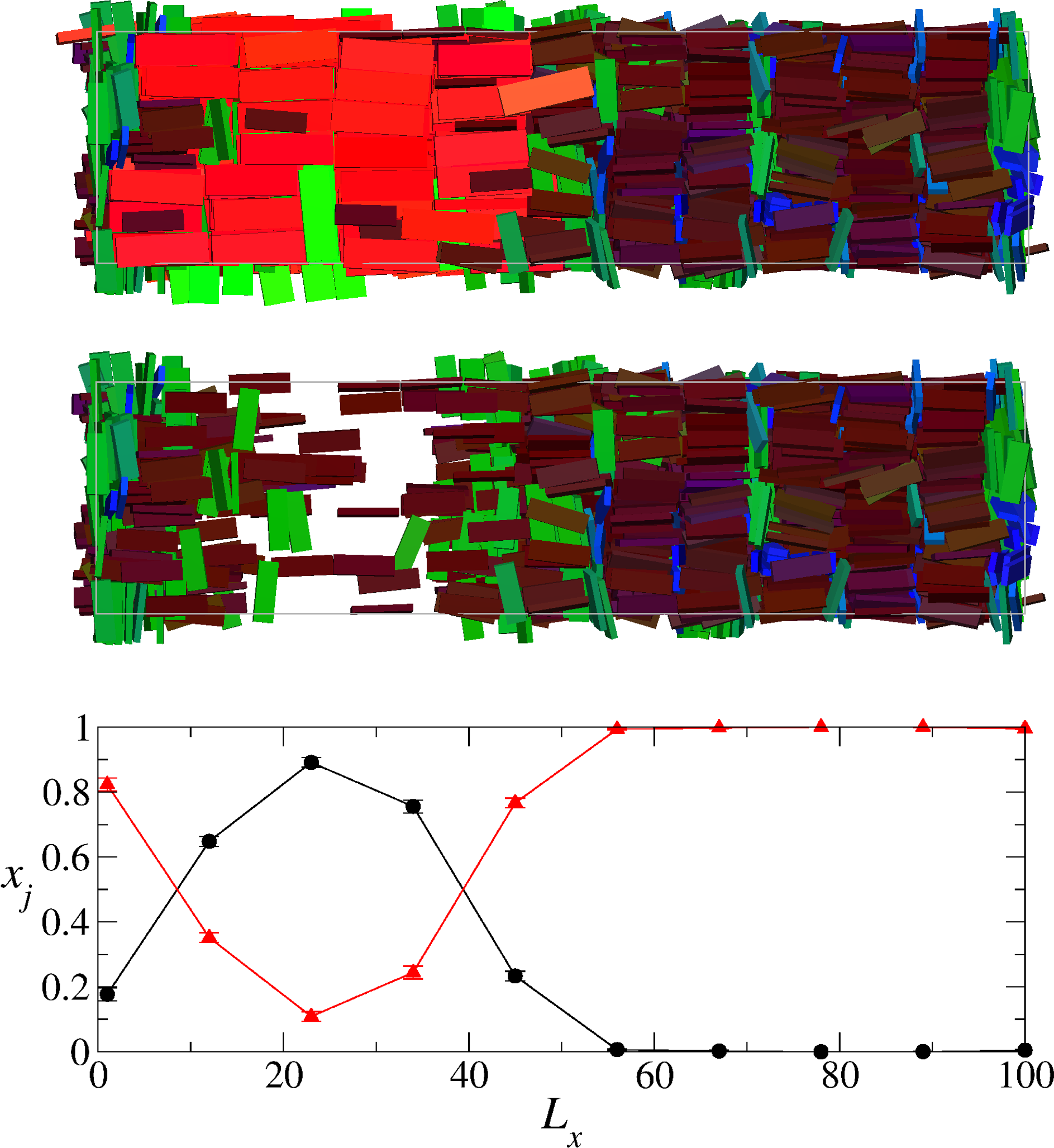}
  \caption{Snapshots of two coexisting smectic phases, one rich in short HBPs and the other rich in long HBPs. Pressure, total number of particles, global composition and size-dispersity index are, respectively, $P^*=0.35$, $N=2000$ and $x_1=0.2$. The top configuration includes long and short HBPs, while the bottom configuration only the short HBPs. The graph reports the concentration profile of long (black circles) and short (red triangles) HBPs along the longest box dimension. Error bars in the bottom frame indicate standard deviations and solid lines are guides to the eye.}
\end{figure}

Typical uniaxial and biaxial order parameters are shown in Figure 7 for a binary mixture at $x_1=0.4$ and $0.1\leq P^* \leq 0.32$. For the sake of clarity, it is important to note that both sets of parameters have been calculated separately for each species and thus provide a measure of the order resulting from the orientation of the HBPs regardless of the phase they belong to. Nevertheless, since the phases observed in the biphasic region predominantly consists of either short or long HBPs, to a very good approximation $S_2$ and $B_2$ can also be employed to measure their orientational order. In particular, the left frame of Figure 7 shows the uniaxial order parameter of the long HBPs (black circles) abruptly increasing at $P^*>0.125$, where the transition from the I to the two-phase region is observed. By contrast, short HBPs (red squares) are very weakly ordered and persist in the I phase up to relatively large pressures. Above $P^*=0.25$, these particles start to form small oriented clusters in the I phase, which can slightly enhance the value of $S_{2,L}$. Additionally, similarly to the organisation observed in the bulk of the Sm$_1$ phase in Figure 4, some of them succeed to diffuse through the layers of the Sm phase and contribute to further increase their global long-range orientational order. As far as the biaxial order parameter $B_2$ is concerned, we notice that long HBPs do show evidence of an appreciable degree of biaxiality at $P^*>0.25$ (right frame of Figure 7). Biaxial Sm phases had also been observed in monodisperse systems of slightly oblate HBPs \cite{cuetos1}. By contrast, no significant biaxiality arises from short HBPs.

\begin{figure}[ht!]
\centering
  \includegraphics[width=0.7\columnwidth]{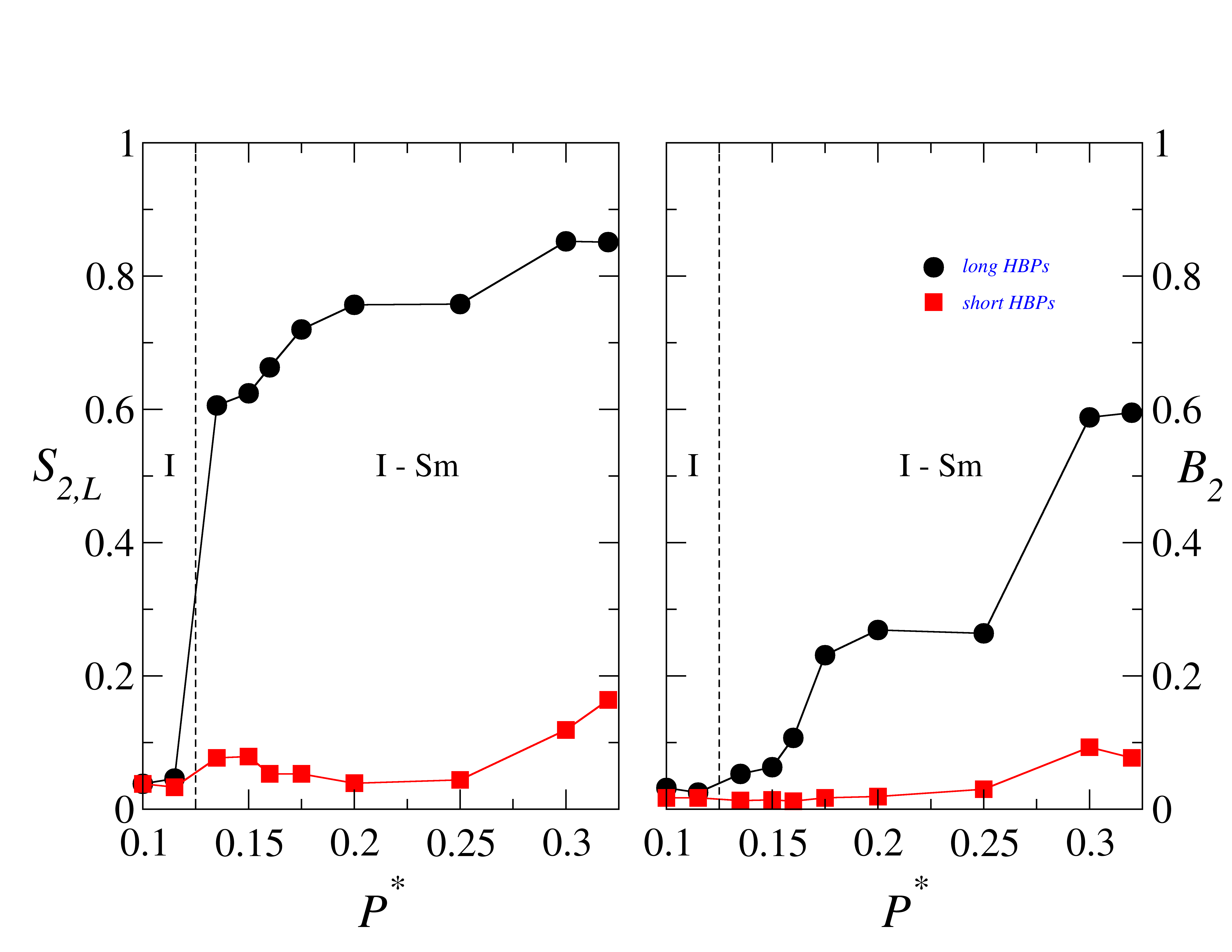}
  \caption{Uniaxial (left) and biaxial (right) order parameters calculated for a binary mixtures of HBPs at $x_1=0.4$ and different values of $P^*$. Vertical dashed lines indicate the transition from the monophasic to the biphasic region. Circles and squares represent, respectively, the order parameter of long and short HBPs in the system.}
\end{figure}

\section{Conclusions}
In summary, we have performed Monte Carlo simulations to investigate the phase behaviour of binary mixtures of short and long HBPs. Upon increasing density, these systems experience a phase transition from a pure I phase, where the two species are homogeneously mixed, to a two-phase region where I phases of mostly short HBPs coexist with Sm phases of mostly long HBPs. At relatively larger densities, a separation between a long-HBP-rich Sm phase (Sm$_1$) and a short-HBP-rich Sm phase (Sm$_2$) is observed. The positionally ordered Sm$_1$ phase incorporates short HBPs within its layers, aligned along the nematic director, and in the inter-layer spacing, aligned perpendicularly to the nematic director. Transverse interlayer particles had been previously observed in Sm phases of monodisperse spherocylinders, although their occurrence was rather rare and usually restricted to one or two transverse particles per interlayer spacing \cite{vanroij, allen1, allen2, Patti1, Patti2}. Although we have not assessed this specific phenomenon quantitatively, here we have detected a significant number of interlayer HBPs oriented perpendicularly to the nematic director, as our pair correlation functions highlight. This intriguing organisation unveils oblate-like smectic layers embedded in a prolate-like smectic phase of mainly long HBPs. By contrast, the Sm$_2$ phase is mostly made of short HBPs which are preferentially ordered along the nematic director. Interestingly, and perhaps surprisingly, our HBPs do not form N$_\text{U}$ phases, even at $x_1=0$ and $x_1=1$. This is in agreement with our recent results on monodisperse HBPs, whose phase diagram at $W\approx\sqrt{LT}$, the geometry explored here, eventually shows a very small region of stability for the N$_\text{U}$ phase. Finally, we do not find any significant evidence of the formation of the N$_\text{B}$ phase, as predicted by Onsager-type theory within the Zwanzig model [13] and by variational cluster expansion theory applied to HBPs with their long axes fully oriented [14]. We conclude that restricting the orientational degrees of freedom by imposing a full or partial alignment of particles can have a significant impact on the extension of the stability region of the N$_\text{B}$ phase. We are aware that the complex free-energy landscape of these systems, which is not fully captured by Onsager's or variational cluster theories, might also play an important role. The experimental observation of especially stable N$_\text{B}$ phases were performed on colloidal dispersions of board-like goethite particles with $L/W \simeq W/T$, exactly as in the present work, but a size polydispersity of 20\%-25\% in all directions was applied [8]. This significant size-dispersity, as compared to the bi-dispersity studied in this work, is most probably a strategic ingredient to circumvent phase separation and destabilise the Sm phase in favour of the N$_\text{B}$ phase. Preliminary simulation results in our group indicate that this is the case, but further investigation is currently ongoing.

\section*{Acknowledgements}

AP acknowledges financial support from EPSRC under grant agreement EP/N02690X/1 and a Researcher Mobility Fellowship awarded by the Royal Society of Chemistry for funding his research stay at the Department of Physical, Chemical and Natural Systems, Pablo de Olavide University. AC acknowledges project CTQ2012-32345 funded by the Junta de Andaluc\'ia-FEDER and C3UPO for HPC facilities provided. AP also acknowledges the assistance given by IT Services and the use of the Computational Shared Facility at the University of Manchester. Finally, we would like to thank Carlos Avenda\~{n}o (University of Manchester) for helpful discussions and his assistance with Figure 1.

\end{document}